\documentclass[reprint,superscriptaddress, amsmath,amssymb,aps]{revtex4-2}

\usepackage{graphicx}
\usepackage{dcolumn}
\usepackage{bm}
\usepackage{gensymb}
\usepackage{braket}
\usepackage{xcolor}
\usepackage[colorlinks=true,citecolor=blue,linkcolor=blue,urlcolor=blue]{hyperref}

\begin{document}

\title{From $\mathrm{RuP_3SiO_{11}}$ to Ideal Geometry: Parallel-Face Octahedra for Kitaev-Dominated Interactions}

\author{Do Hoon Kiem}
\email{kimd4@ornl.gov}
\affiliation{Materials Science and Technology Division, Oak Ridge National Laboratory, Oak Ridge, Tennessee 37831, USA}

\author{Satoshi Okamoto}
\email{okapon@ornl.gov}
\affiliation{Materials Science and Technology Division, Oak Ridge National Laboratory, Oak Ridge, Tennessee 37831, USA}

\author{Tom Berlijn}
\email{berlijnt@ornl.gov}
\affiliation{Center for Nanophase Materials Sciences, Oak Ridge National Laboratory, Oak Ridge, Tennessee 37831, USA}

\date{\today}

\begin{abstract}
$\mathrm{RuP_3SiO_{11}}$ has recently been identified as a Kitaev spin liquid candidate via neutron scattering. To assess its proximity to the Kitaev limit, we perform \textit{ab initio} calculations of magnetic interactions, finding good agreement for the Kitaev and off-diagonal exchanges, while the Heisenberg exchange is ferromagnetic rather than the experimentally fitted antiferromagnetic one. We also identify a sizable antiferromagnetic out-of-plane exchange not captured in previous analyses. Extending the study to eight substitutional analogs in the $MT_{3}Q\mathrm{O}_{11}$ family ($M$ = Ru, Os; $T$ = P, As; $Q$ = Si, Ge), we observe strong variations in magnetic interactions, highlighting the tunability of this materials class. Finally, inspired by $\mathrm{RuP_3SiO_{11}}$, which exhibits a near parallel-face octahedral arrangement, we compare idealized geometries and show that the parallel-face configuration yields a Kitaev-dominated regime, suggesting a new structural route to the Kitaev limit.
\end{abstract}

\maketitle

\begin{figure}
    \centering
    \includegraphics[width=1.0\linewidth]{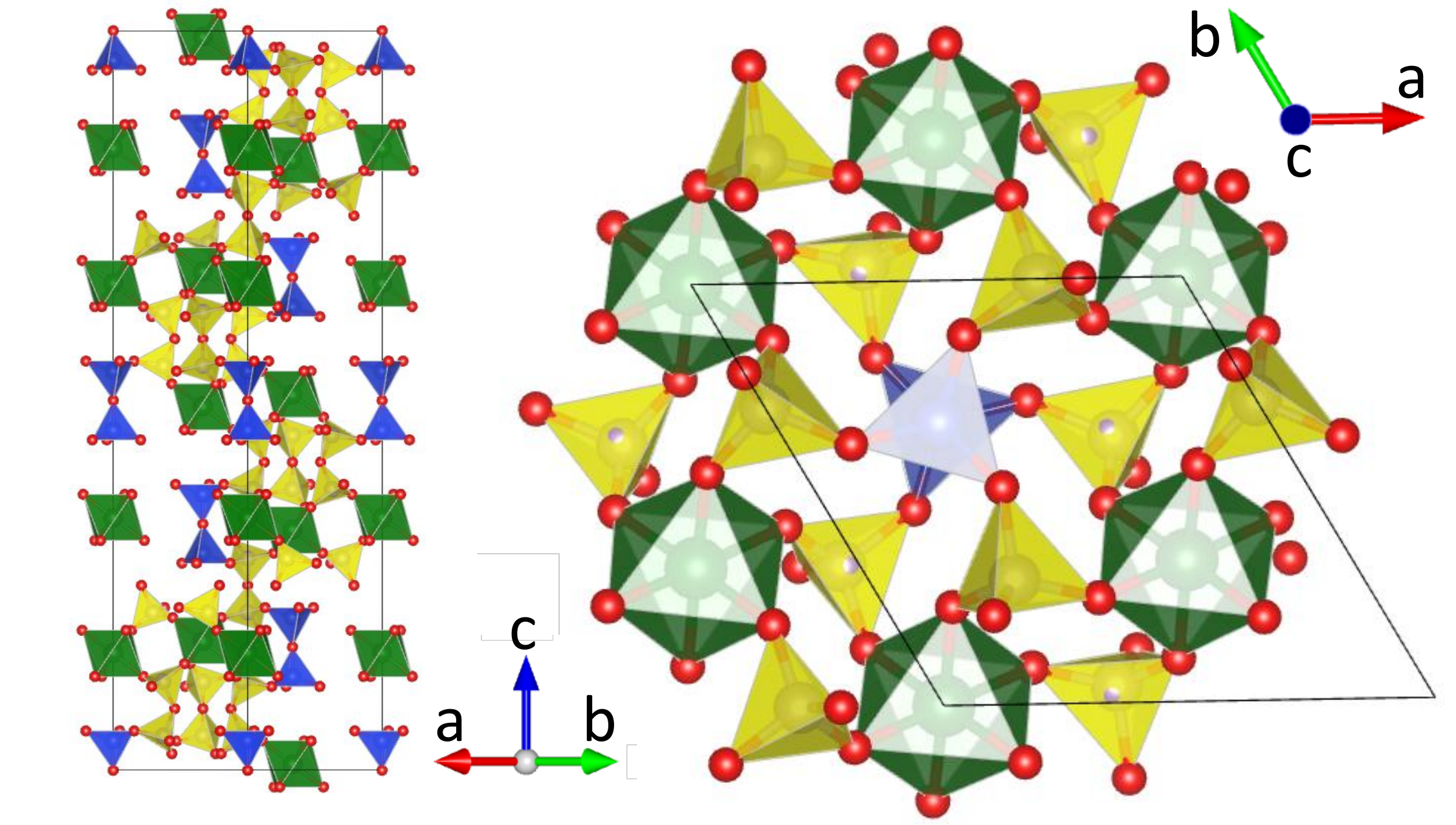}
    \caption{Structure of the Kitaev candidate $\mathrm{RuP_3SiO_{11}}$. Left: side view of stacked honeycomb layers. Right: top view of a honeycomb layer, with green octahedra representing RuO$_6$, and yellow and blue bitetrahedra representing pyrophosphate (P$_2$O$_7$) and pyrosilicate (Si$_2$O$_7$), respectively.}
    \label{fig1}
\end{figure}

The Kitaev model predicts that a honeycomb lattice with bond--dependent Ising interactions can host a quantum spin liquid state characterized by fractionalized Majorana fermions, with exciting potential applications in quantum information science \cite{kitaev2006anyons, freedman2003topological}. A major breakthrough was the realization that the Kitaev model could emerge in materials where crystal field effects, strong spin-orbit coupling (SOC), and electron correlations give rise to effective spin-1/2 moments \cite{jackeli2009mott, takagi2019concept,chaloupka2010kitaev}. Since then, several Kitaev candidate materials have been explored, including honeycomb iridates ($A_2$IrO$_3$; $A$=Na, Li) \cite{singh2012relevance, gretarsson2013crystal}, $\alpha$-$\mathrm{RuCl}_3$ \cite{kim2015kitaev, do2017majorana, banerjee2016proximate}, and cobaltates \cite{kim2021antiferromagnetic, songvilay2020kitaev, sanders2022dominant, Motome_2020}. Promising signatures of Kitaev physics have been observed in $\alpha$-$\mathrm{RuCl}_3$, such as half-quantized thermal Hall effect \cite{yokoi2021half, kasahara2018majorana}, field-induced spin gap \cite{hentrich2018unusual}, and recent device-based studies \cite{klocke2022thermal, zhang2025evidence, zhang2024anisotropic}, though the true quantum spin liquid nature of $\alpha$-$\mathrm{RuCl}_3$ remains unclear. In particular, first-principles calculations and experimental fits \cite{kim2015kitaev, winter2016challenges,eichstaedt2019deriving,HYKee2023Co,Winter_2022,ran_2017,wu_2018} have revealed the presence of competing interactions—including first- and third-neighbor Heisenberg exchanges and significant anisotropic $\Gamma$ interactions—which may drive these systems away from the ideal Kitaev limit.

While the search for Kitaev spin liquids has traditionally focused on materials with edge-sharing octahedra, candidate materials featuring separated octahedra have received much less attention. In these systems, the magnetic octahedra are spaced apart either by metallic-organic frameworks (MOFs) \cite{yamada2017designing, yamada2017crystalline, zhang2024two, burzuri2023quantum}, or by inorganic spacer units~\cite{abdeldaim2024kitaev,liu2023ba9re2}. The material studied in this work, $\mathrm{RuP_3SiO_{11}}$, provides an example of the latter case. As shown in Fig.~\ref{fig1}, the RuO$_6$ octahedra form a honeycomb lattice within the plane, with neighboring octahedra spatially separated by two pyrophosphate (P$_2$O$_7$) spacer units, while a pyrosilicate (Si$_2$O$_7$) group resides at the center of each honeycomb hexagon. The pyrophosphate spacers further connect the honeycomb layers along the out-of-plane direction, resulting in coupled two-dimensional magnetic layers. To date, only a handful of separated octahedra compounds have been explored, including $\textrm{RuP}_3\textrm{SiO}_{11}$ \cite{fukuoka1996synthesis,abdeldaim2024kitaev}, Ru(C$_2$O$_4$)$_3$ \cite{yamada2017designing}, Ru(C$_4$N$_6$H$_4$)$_3$ \cite{yamada2017designing}, and Ba$_9$RE$_2$(SiO$_4$)$_6$ \cite{liu2023ba9re2}. Notably, for Ba$_9$RE$_2$(SiO$_4$)$_6$, no magnetic ordering has been observed down to 0.15 K. While these spaced Kitaev materials often exhibit greater structural complexity, this very complexity also dramatically expands the combinatorial landscape, offering a vast and largely untapped space of potential Kitaev spin liquid candidates.

One drawback of separating the octahedra is a reduction in the strength of magnetic interactions, which can make experimental detection more challenging. However, recent inelastic neutron scattering (INS) experiments on RuP$_3$SiO$_{11}$ have demonstrated that the magnetic interactions in this system remain within the realm of experimental observability, enabling the direct identification of significant Kitaev interactions \cite{abdeldaim2024kitaev}. A key advantage of spaced octahedra however, is the suppression of second- and third-nearest-neighbor in-plane magnetic exchanges. In particular, the third-nearest-neighbor Heisenberg interaction is known to stabilize competing antiferromagnetic zigzag order in edge-sharing systems \cite{Garlea2024ZigzagReview,Kimchi2011afmZZ,songvilay2020kitaev,lin2021field,hwan2015direct}, so its reduction helps to promote the Kitaev spin liquid state. From a theoretical perspective, increased separation between octahedra also brings these materials closer to the strong-coupling regime, which is commonly assumed in perturbative derivations of magnetic interactions, as employed in this work. Finally, reducing the in-plane Kitaev interactions through increased octahedral separation can enhance the relative strength of out-of-plane interactions. This feature may be important for the realization of gate-controlled anyon detection devices \cite{halasz2024gate}, opening new avenues for quantum information applications.

In this Letter, we present first-principles magnetic exchange interactions for the Kitaev candidate material $\mathrm{RuP}_3\mathrm{SiO}_{11}$. Using density functional theory and perturbation theory, we find reasonable agreement with recent experimental fits, though our results identify antiferromagnetic exchange and strong interlayer couplings. We also find that substitutions with Os, As, and Ge can greatly vary these magnetic interactions. Most importantly, we note that $\mathrm{RuP}_3\mathrm{SiO}_{11}$ is characterized by separated octahedra with parallel faces. To understand the physical consequences of this geometry, we compare idealized models of edge-sharing, parallel-edge, and parallel-face octahedra. We find that the parallel-face configuration uniquely suppresses competing interactions, through geometrically controlled orbital alignment that reshapes the hopping hierarchy, suggesting a new structural strategy for the design of Kitaev materials.

\begin{figure}
    \centering
    \includegraphics[width=1.0\linewidth]{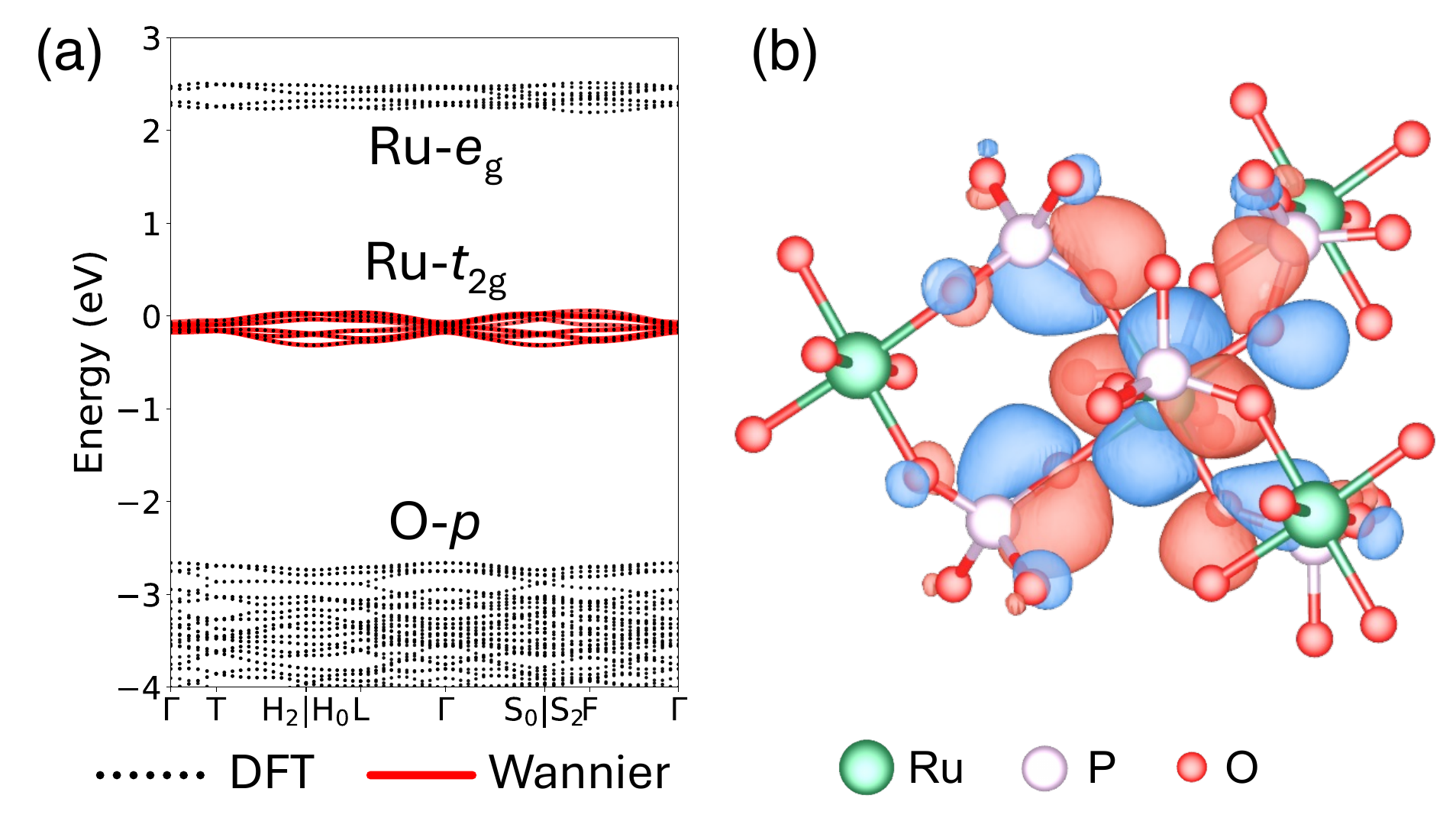}
    \caption{(a) The $\mathrm{RuP_{3}SiO_{11}}$ band structure from density functional theory (black dots) and Wannier-function calculations (red lines). (b) A corresponding Ru-$d_{xz}$ Wannier function. An isosurface value of 1.0 electrons/\AA$^{3}$ is used.}
    \label{fig2}
\end{figure}

To derive the magnetic exchange interactions in $MT_{3}Q\mathrm{O}_{11}$ compounds and the idealized model structures, we employ a standard three-step procedure. First, we perform density functional theory (DFT) calculations~\cite{kresse1996efficient,blochl1994projector,perdew1996generalized,bellaiche2000virtual}. Next, we construct a low-energy tight-binding Hamiltonian by projecting onto Ru and Os $t_{2g}$ Wannier functions~\cite{pizzi2020wannier90,zhi2022wannsymm,momma2011vesta}. A comparison between the DFT and Wannier-interpolated band structures, along with one representative Ru-$xz$ Wannier orbital, is shown in Fig.~\ref{fig2}. Input and output files for the DFT and Wannier calculations are available at~\cite{data_repository}. In the final step, we extract the exchange couplings via strong-coupling perturbation theory, treating the intersite hopping as a perturbation~\cite{kim2015kitaev, winter2016challenges,eichstaedt2019deriving,HYKee2023Co,Rau2018Yb,Winter_2022,ChengPeng2024}. The local Hamiltonian includes the trigonal crystal field, atomic spin–orbit coupling (SOC), and local interactions in the Kanamori form. The SOC is included as an atomic $\lambda\,\mathbf{L}\cdot\mathbf{S}$ term, where $\lambda$ is extracted from the DFT Wannier Hamiltonian. The local interactions are parametrized with intra-orbital Hubbard $U=3$~eV, inter-orbital $U'=2$~eV, and Hund's coupling $J_{\mathrm{H}}=0.5$~eV. The resulting spin Hamiltonian can be written in compact form as
$H = \sum_{n,\gamma}\sum_{\langle ij\rangle_n^\gamma} 
\mathbf{S}_i \cdot \mathcal{J}_n^\gamma \cdot \mathbf{S}_j$, 
where $n = 1,2,3,\mathrm{out}$ labels the neighbor shell (first, second, and third in-plane neighbors, and nearest out-of-plane bonds), and $\gamma = X,Y,Z$ denotes the bond direction. The exchange matrix for the $Z$-bond is parameterized as
\begin{align}
\mathcal{J}_n^{Z} =
\begin{pmatrix}
J_n + \xi_n & \Gamma_n + D_n^z & \Gamma'_n + \zeta_n - D_n^y \\
\Gamma_n - D_n^z & J_n - \xi_n & \Gamma'_n - \zeta_n + D_n^x \\
\Gamma'_n + \zeta_n - D_n^y & \Gamma'_n - \zeta_n - D_n^x & J_n + K_n
\end{pmatrix}.
\end{align}
The matrices $\mathcal{J}_n^{X}$ and $\mathcal{J}_n^{Y}$ are obtained from $\mathcal{J}_n^{Z}$ via $C_3$ cyclic permutation of spin components. The Dzyaloshinskii--Moriya interactions vanish for first- and third-nearest in-plane bonds ($n=1,3$). For the first-nearest out-of-plane bond ($n=\mathrm{out}$), there is only a single orientation (no $X,Y,Z$ distinction), corresponding to bonds perpendicular to the plane, and the parameters satisfy $\Gamma'_{\mathrm{out}} = \Gamma_{\mathrm{out}}$, $\xi_{\mathrm{out}} = \zeta_{\mathrm{out}}=0$ and $D_n^x = D_n^y = D_n^z \equiv D_{\mathrm{out}}$. Further computational details are provided in the Supplemental Material.

\begin{table}[h]
\centering
\caption{Maximum elements of the in-plane exchange matrices $\mathcal{J}_n$ for $\mathrm{RuP_{3}SiO_{11}}$.}
\label{Tab1}
\begin{tabular}{lcr}
\hline
Neighbor & Symbol & Value ($\mu$eV) \\ \hline
First  & $\max(\mathcal{J}_1) = \Gamma_1$ & $502.5$   \\
Second & $\max(\mathcal{J}_2) = \xi_2$    & $-9.0$   \\
Third  & $\max(\mathcal{J}_3) = \zeta_3$    & $0.370$  \\ \hline
\end{tabular}
\end{table}

Our first result is that the in-plane long-range interactions beyond nearest neighbors are strongly suppressed, exhibiting an approximately exponential decay. In Table~\ref{Tab1}, we focus on the dominant parameter for each in-plane exchange matrix. For $\mathcal{J}_1$, this is the anisotropic interaction $\Gamma_1$; for $\mathcal{J}_2$, it is $\xi_2$; and for $\mathcal{J}_3$, it is $\zeta_3$. Quantitatively, the largest second-nearest-neighbor coupling is roughly 50 times smaller than the corresponding nearest-neighbor term, and the third-nearest-neighbor interaction is further suppressed by an additional factor of $\sim 20$. These results indicate that spatially separated octahedra effectively suppress unwanted long-range interactions, which in edge-sharing Kitaev candidates are known to give rise to competing magnetic phases such as the zigzag antiferromagnetic state.

\begin{table}[h]
\centering
\caption{Comparison of the largest exchange interactions (meV) for Ru- and Os-based compounds. Only the largest elements of each exchange matrix are shown. Experimental values for RPSO correspond to the best-fit solution~\cite{abdeldaim2024kitaev}. Abbreviations: RPSO = $\mathrm{RuP_3SiO_{11}}$, OAGO = $\mathrm{OsAs_3GeO_{11}}$.}
\label{Tab2}
\begin{ruledtabular}
\begin{tabular}{lrrr}
Parameter & RPSO (exp) & RPSO (calc) & OAGO (calc) \\
\hline
$J_1$       & $0.32$   & $-0.23$ & $-0.27$  \\
$K_1$       & $-0.54$  & $-0.25$ & $2.13$   \\
$\Gamma_1$  & $0.44$   & $0.50$  & $1.99$   \\
$\Gamma'_1$ & $-0.12$  & $-0.25$ & $-0.05$  \\
$J_{\mathrm{out}}$      & --       & $0.42$  & $3.87$   \\
$\Gamma_{\mathrm{out}}$ & --       & $0.07$  & $2.79$   \\
$D_{\mathrm{out}}$      & --       & $0.10$  & $-5.34$  \\
\end{tabular}
\end{ruledtabular}
\end{table}

We next compare our computed exchange interactions for $\mathrm{RuP_3SiO_{11}}$ with the best-fit parameters extracted from recent inelastic neutron scattering~\cite{abdeldaim2024kitaev} (Table~\ref{Tab2}). In particular, $K_1$, $\Gamma_1$, and $\Gamma_1'$ all have the same sign and are of comparable magnitude. This represents reasonably good agreement with experiment, given that the results are obtained without parameter tuning and that the interactions themselves are weak. There are, however, two notable differences between theory and experiment. First, we find a ferromagnetic nearest-neighbor Heisenberg $J_1$ instead of the antiferromagnetic value obtained in the experimental fit. Second, we predict sizable antiferromagnetic out-of-plane interactions mediated by the pyrophosphate groups, which consist of two tetrahedra sharing an oxygen atom (colored in yellow in Fig.~\ref{fig1}). These out-of-plane couplings were not included in the previous experimental analysis, but pose a significant obstacle to realize the Kitaev spin-liquid state in this material.

Turning to chemical substitutions, we considered replacing Ru with Os, P with As, and Si with Ge, resulting in a total of eight compounds. The main conclusion is that these substitutions cause significant changes in the magnetic interactions. The lattice parameters and internal atomic positions for all $MT_3Q\mathrm{O}_{11}$ (M = Ru, Os; T = P, As; Q = Si, Ge) compounds were optimized with a force criterion of 0.001 eV/Å in the absence of SOC. We also computed the exchange interactions for the experimental structure of $\mathrm{RuP_3SiO_{11}}$ from Ref.~\cite{abdeldaim2024kitaev} and found only minor differences with the relaxed simulation. Full results for all parameters for all eight compounds are provided in the Supplemental Material. In Table~\ref{Tab2}, we present the dominant interactions for $\mathrm{OsAs_3GeO_{11}}$ for comparison with $\mathrm{RuP_3SiO_{11}}$. From this comparison, we see that the substitutions reverse the Kitaev interaction from ferromagnetic to antiferromagnetic and increase its magnitude by roughly an order of magnitude, while the competing nearest in-plane Heisenberg interactions $J_1$ and $\Gamma_1'$ remain about one or two orders of magnitude smaller. Unfortunately, the $\Gamma_1$ and the nearest out-of-plane Heisenberg interaction $J_{\mathrm{out}}$ are also strongly enhanced by these chemical substitutions and remain sizable relative to the Kitaev term in the cases considered. Full sets of parameters for the first- and second-nearest-neighbor in-plane interactions, as well as the nearest out-of-plane couplings, for all eight substitutional compounds are provided in the Supplemental Material.

\begin{figure}
    \centering
    \includegraphics[width=1.0\linewidth]{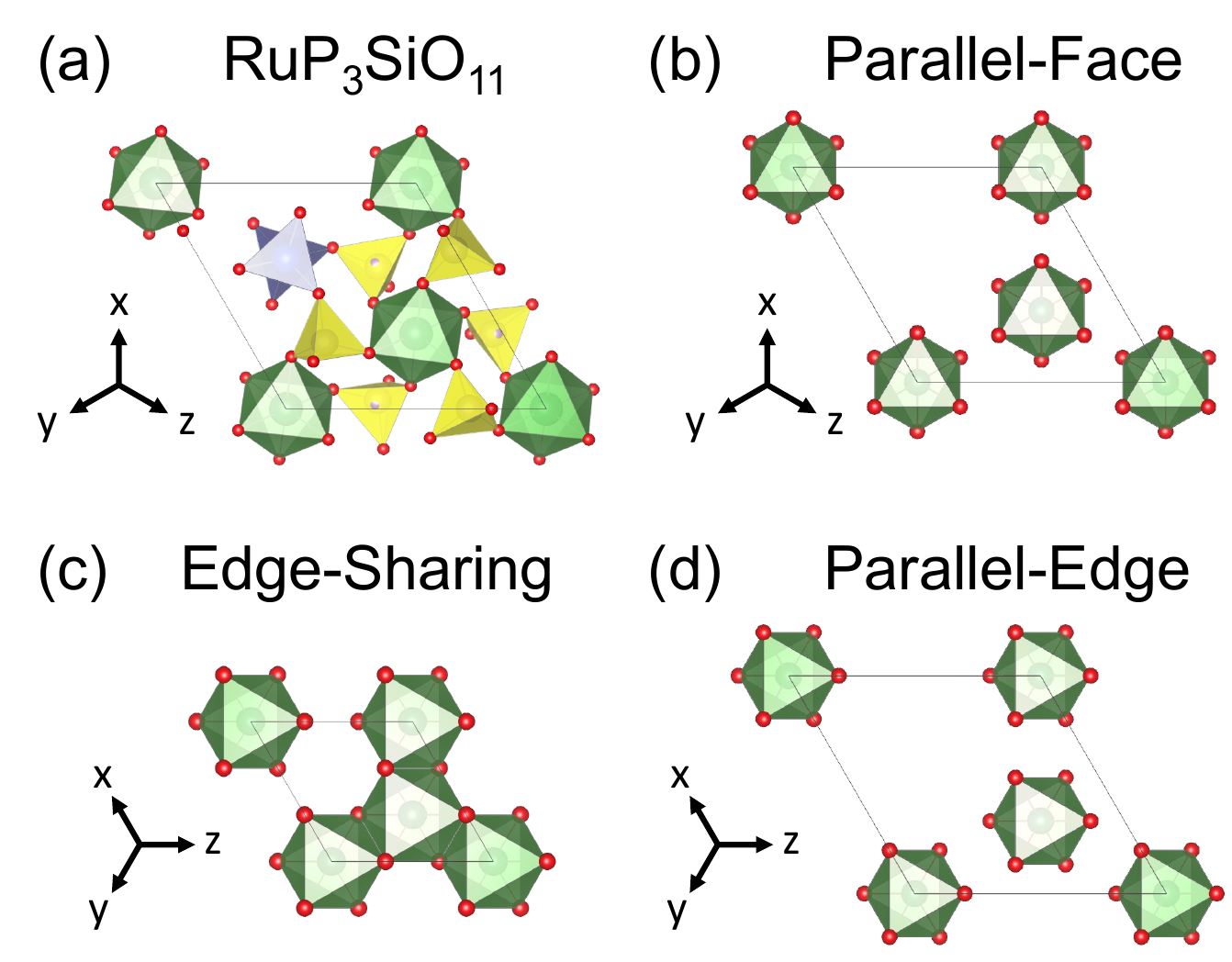}
    \caption{(a) $\mathrm{RuP_{3}SiO_{11}}$ octahedra. Idealized models of octahedra with (b) parallel faces, (c) shared edges, and (d) parallel edges.}
    \label{fig3}
\end{figure}

Finally, we identify an important aspect of the ruthenium phosphosilicate structure: the octahedra exhibit nearly parallel faces (Fig.~\ref{fig3}). While Kitaev candidates with edge-sharing octahedra or octahedra with parallel edges have been widely considered~\cite{TREBST}, the case of parallel faces has, to our knowledge, not been explicitly identified or discussed. To assess the general significance of this geometry, we performed idealized model calculations in which all atoms except those forming perfectly undistorted octahedra were removed. We then considered three configurations: edge-sharing, parallel-edge, and parallel-face arrangements. For the edge-sharing case, we used the in-plane lattice parameter $a = 5.973$~\AA\ corresponding to $\alpha-\mathrm{RuCl_3}$ from Ref.~\cite{Mu}. For the parallel-edge and parallel-face configurations, we used $a = 10.34555$~\AA. As will be discussed below, this choice allows for the clearest explanation of the regime of dominant Kitaev interactions, though similar behavior occurs for nearby lattice spacings. In all cases, the ligand positions were adjusted to yield ideal octahedra with $90^\circ$ Cl--Ru--Cl bond angles. To ensure a $\mathrm{Ru}^{3+}$ electronic configuration in the parallel-edge and parallel-face models, we employed a virtual crystal approximation by replacing the ligand pseudopotentials with a 50\% mixture of Cl and Ar. This results in well-isolated Ru-$t_{2g}$ bands (shown in the Supplemental Material), which are similar to those in Fig.~\ref{fig2}(a) for $\mathrm{RuP_3SiO_{11}}$.

\begin{table}[htbp]
\centering
\caption{Electronic hopping integrals (meV) and magnetic exchange parameters ($\mu$eV) for idealized $\mathrm{Ru}^{3+}$ geometries.}
\label{Tab3}
\begin{ruledtabular}
\begin{tabular}{lrrr}
Parameter & Edge-Sharing & Parallel-Edge & Parallel-Face \\
\hline
Magnetic ($\mu$eV) & & & \\
$J$             & $296.7$   & $5763.3$  & $14.6$     \\
$K$             & $-2061.3$ & $2822.3$  & $-352.1$   \\
$\Gamma$        & $3780.0$  & $-907.5$  & $-21.4$    \\
$\Gamma'$       & $-922.0$  & $68.7$    & $-18.3$    \\
 \\
\hline
Electronic (meV) & & & \\
$\Delta$     & $-22.9$  & $5.4$   & $7.7$     \\
$t_{xy-xy}$     & $-215.1$  & $-179.6$   & $-0.2$     \\
$t_{xy-xz}$     & $-4.8$    & $2.8$      & $4.5$      \\
$t_{yz-yz}$     & $62.5$    & $-1.1$     & $6.0$     \\
$t_{yz-xz}$     & $115.8$   & $-31.4$    & $-35.0$    \\
\end{tabular}
\end{ruledtabular}
\end{table}

Remarkably, the parallel-face configuration yields a nearly ideal Kitaev model. The corresponding magnetic and electronic parameters are summarized in Table~\ref{Tab3}. Focusing first on the magnetic interactions, we find that in the parallel-face case the competing Heisenberg and $\Gamma$ interactions are more than an order of magnitude smaller than the Kitaev term, while $\Gamma'$ is two orders of magnitude smaller. This behavior is not observed in the other geometries. In the edge-sharing configuration, although the Kitaev interaction exceeds the Heisenberg term, it remains smaller than $\Gamma$ by roughly a factor of two. In contrast, for the parallel-edge configuration, the relative hierarchy is reversed. Only in the parallel-face geometry does the Kitaev interaction clearly dominate over all competing terms. This demonstrates that the orientation of separated octahedra is a powerful tuning parameter for approaching the Kitaev limit.

\begin{figure}[htbp]
    \centering
    \includegraphics[width=1.0\linewidth]{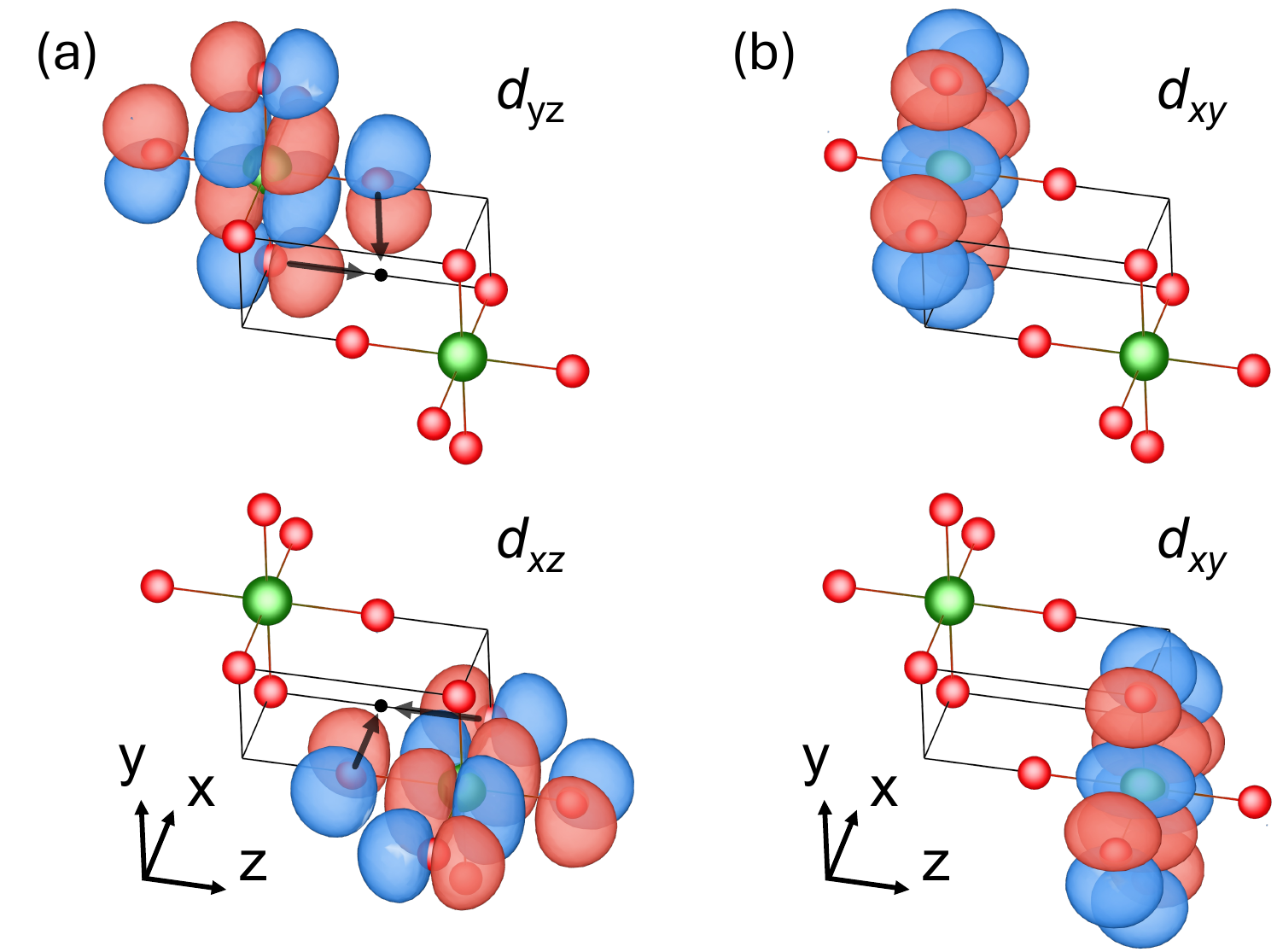}
		\caption{Wannier functions for the parallel-face configuration illustrating the geometric origin of the hopping hierarchy for the nearest Z-bond. (a) Ru-$xz$ and Ru-$yz$ Wannier functions exhibit strong overlap as their ligand tails point toward a common point on the edge of the rectangular prism defined by the Ru ions and bridging ligands. (b) Ru-$xy$ Wannier functions lie in the planes of the rectangular prism separated along its long edge (the $z$ direction) and show minimal overlap, leading to suppressed $xy$–$xy$ hopping. An isosurface value of 1.42 electrons/\AA$^{3}$ is used.}
    \label{fig4}
\end{figure}

To better understand the origin of the nearly ideal Kitaev limit in the parallel-face configuration, we next examine the electronic model parameters listed in Table~\ref{Tab3}. Specifically, we focus on the nearest-neighbor $Z$-bond hopping integrals. A striking difference becomes immediately apparent. Whereas in the edge-sharing and parallel-edge configurations the Ru-$xy$ to Ru-$xy$ hopping dominates, in the parallel-face configuration this diagonal hopping is reduced to nearly zero. The indirect Ru-$xz$–Ru-$yz$ hopping, subleading in the edge-sharing and parallel-edge configurations, instead becomes dominant in the parallel-face configuration. This is the key feature that drives parallel-face configuration close to the Kitaev limit.

To clarify the hierarchy of hopping amplitudes, we examine the Wannier functions shown in Fig.~\ref{fig4} for the parallel-face configuration. A key structural difference is that the neighboring Ru ions and bridging ligands lie in a common plane in the edge-sharing and parallel-edge cases, whereas in the parallel-face configuration they form a three-dimensional arrangement. In particular, they define a rectangular prism with the long axis along the $z$ direction and the Ru ions located on opposite diagonals. This geometry directly controls the overlap of the Wannier functions. As shown in Fig.~\ref{fig4}(a), the ligand tails of the Ru-$xz$ and Ru-$yz$ Wannier functions point toward a common point on one of the long edges of this prism, resulting in maximal overlap and dominant $xz$–$yz$ hopping. In contrast, the Ru-$xy$ Wannier functions in Fig.~\ref{fig4}(b) lie primarily within planes perpendicular to the long edge of the rectangular prism  and therefore exhibit minimal overlap, leading to strongly suppressed $xy$–$xy$ hopping. Intermediate cases, such as $xy$–$xz$ and $xz$–$xz$, show partial alignment: the former involves only one orbital with a $z$ component, while in the latter both orbitals have $z$ character but do not point toward a common region (see Fig.~S3), resulting in subleading hopping amplitudes. This hierarchy contrasts with the edge-sharing and parallel-edge configurations, where the relevant orbitals and ligands lie in a common plane, enabling a dominant $xy$–$xy$ hopping either through ligand-mediated processes or direct $d$–$d$ contributions (see Fig.~S3). In the parallel-face configurations considered here, we used $a = 10.34555~\text{\AA}$ such that the Ru–ligand bond length matches the in-plane projection of the Ru–Ru separation, enabling the rectangular prism construction illustrated in Fig.~\ref{fig4}. However, the resulting dominance of $xz$–$yz$ hopping, and thus of the Kitaev interaction, is not fine-tuned. As shown in the Supplemental Material, we also explored $a \sim 10~\text{\AA}$ and $a=11~\text{\AA}$ and find that this hierarchy is robust and evolves systematically with increased lattice spacing. We further find that varying the trigonal crystal field modifies the magnetic parameters but leaves the Kitaev-dominant character intact.

In conclusion, we have identified the parallel-face orientation of octahedra as a new route toward realizing the Kitaev limit in materials. While existing compounds do not yet fully realize this ideal geometry, $\mathrm{RuP_3SiO_{11}}$ already captures its essential feature, with its octahedra being close to the parallel-face configuration. The remaining requirements are structural. In our analysis, bonding angles are taken to be $90^\circ$, although the sensitivity to deviations from this value remains to be established. In addition, spacers, such as the $\mathrm{P_2O_7}$ units in $\mathrm{RuP_3SiO_{11}}$, must be controlled to avoid mediating out-of-plane couplings and disrupting the geometric orbital alignment that underlies the hopping hierarchy. Despite these structural constraints, the identification of a new pathway for realizing the Kitaev limit is significant, especially as Kitaev candidates based on edge-sharing octahedra appear extensively explored and most compounds exhibit significant competing interactions. In contrast, materials with separated octahedra naturally suppress longer-range couplings and offer a vastly expanded combinatorial design space, which can be systematically explored using modern approaches such as generative models and graph neural networks.

\begin{acknowledgments}
We thank Jiaqiang Yan for suggesting the study of the $MT_{3}Q\mathrm{O}_{11}$ family, Colin Sarkis for helpful discussions, and Christian Balz for insightful analysis. This work was supported by the U.S. Department of Energy, Office of Science, National Quantum Information Science Research Centers, Quantum Science Center. This research used resources of the Compute and Data Environment for Science (CADES) at the Oak Ridge National Laboratory, which is supported by the Office of Science of the U.S. Department of Energy under Contract No. DE-AC05-00OR22725. 

\end{acknowledgments}

\bibliographystyle{apsrev4-2}
\bibliography{RPSOreference}

\end{document}